\newcommand{\PRL}[3]{Phys.\ Rev.\ Lett.\ {\bf #1},\ #2 (#3)}

\newcommand{\SC}[3]{Science\ {\bf #1},\ #2 (#3)}

\newcommand{\SOV}[3]{Sov.\ Phys.\ JETP\ {\bf #1},\ #2 (#3)}
\newcommand{\PRA}[3]{Phys.\ Rev.\ A\ {\bf #1},\ #2 (#3)}

\newcommand{\PRE}[3]{Phys.\ Rev.\ E\ {\bf #1},\ #2 (#3)}

\newcommand{\JPA}[3]{J.\ Phys.\ A:\ Math.\ Gen.\ {\bf #1},\ #2 (#3)}

\newcommand{\EPL}[3]{Eur.\ Phys.\ Lett.\ {\bf #1},\ #2 (#3)}

\documentclass[aps,pra,superscriptaddress,secnumarabic,twocolumn,showpacs,
amsmath,amsfonts,amssymb,floatfix]{revtex4}
\usepackage{graphicx}
\usepackage{color}
\usepackage{epsf}
\begin{document}

\preprint{APS/123-QED}

\title{Sinusoidal Excitations in Two Component Bose-Einstein Condensates in a Trap}

\author{Priyam Das}\email{priyam@iiserkol.ac.in}
 \affiliation{Indian Institute of Science Education and Research (IISER), Salt Lake,
Kolkata - 700106, India}
\author{T. Soloman Raju}
\affiliation{Physics Group, Birla Institute of Technology and Science-Pilani, Goa, 403 726, India}
\author{Utpal Roy}\email{utpal.roy@unicam.it}
\affiliation{Department of Physics, Univ. of Camerino, Camerino (MC), Italy}
\author{Prasanta K. Panigrahi}\email{prasanta@prl.res.in}
 \affiliation{Physical Research Laboratory, Navrangpura, Ahmedabad -
380009, India}
 \affiliation{Indian Institute of Science Education and Research (IISER), Salt Lake,
Kolkata - 700106, India}

\begin{abstract}
The non-linear coupled Gross-Pitaevskii equation governing the dynamics of the two component Bose-Einstein condensate (TBEC) is shown to admit sinusoidal, propagating wave solutions in quasi one dimensional geometry in a trap. The solutions exist for a wide parameter range, which illustrates the procedure for coherent control of these modes through temporal modulation of the parameters, like scattering length and oscillator frequency. The  effects of time dependent coupling and the trap variation on the condensate profile are explicated. The TBEC has also been investigated in presence of an optical lattice potential, where the superfluid phase is found to exist under general conditions.
\end{abstract}


\maketitle


Much theoretical work has already gone into studying the ground
state solutions of the coupled Gross-Pitaevskii (GP) equations
describing multi-component BECs \cite{Ho5,Pu5,eberly15,eberly25}.
TBEC has been observed, where the two hyperfine levels of $^{87}Rb$
\cite{Myatt5,Hall5} act as the two components. In this case, a
fortuitous coincidence in the triplet and singlet scattering lengths
has led to the suppression of exoergic spin-exchange collisions,
which lead to heating and resultant loss of atoms. A number of
interesting features, like the preservations of the total density
profile and coherence for a characteristically long time, in the
face of the phase-diffusing couplings to the environment and the
complex relative motions \cite{matthews5}, point to the extremely
interesting dynamics of the TBEC. TBEC has been produced in a system
comprising of $^{41}K$ and $^{87}Rb$, in which sympathetic cooling
of $Rb$ atoms was used to condense the $K$ atoms \cite{Modungo5}. It
has also been observed in $^{7}Li-^{133}Cs$ \cite{Mudrich5} and
$^{87}Rb-^{133}Cs$ systems \cite{haas5}.

The presence of nonlinearities in BECs \cite{pethick5}, make them
ideal candidates for observation of solitary waves, ubiquitous to
non-linear media \cite{Eberly35,das5,soliton5,GPA5}.
In the TBEC, a number of investigations, primarily devoted to the study of
localized solitons, have been carried out recently
\cite{anderson5,nath5,michal5,laksh35,Arthur5}. The coincidence of
singlet-triplet coupling in $^{87}Rb$, leads to the well known
Manakov system \cite{Manakov5} in weak coupling quasi-one
dimensional scenario \cite{jackson5,salasnich5}. The rich dynamics
of solitons in this integrable system has received considerable
attention in the literature \cite{laksh15,laksh25,Lazarides5,
Derevyanko5}. The effect of spatial inhomogeneity, three-dimensional
geometry, and dissipation on TBEC have been examined.
However, the periodic solitary waves have not received much
attention in the literature, particularly in the presence of the harmonic trap \cite{Konotop}.
Periodic sinusoidal excitations are
natural in linear systems. In nonlinear models periodic cnoidal
waves can be present. It is worth mentioning that, in non-linear
resonant atomic media, cnoidal excitations have been experimentally
generated \cite{Salamo5,Shultz5}, where relaxation naturally led to
the atomic level population necessary for the existence of these
nonlinear periodic waves \cite{Panigrahi5}.

Here we analyze the solutions of a generic TBEC model in a quasi-one
dimensional geometry for periodic solutions. Interestingly, we
find exact sinusoidal wave solutions in this system in the presence of
a harmonic trap, which do not occur in the single component case.
The presence of two components leads to these waves, whose energy
difference are controlled by the cross phase modulation (XPM). In
presence of time dependent trap and
scattering length, these waves can be compressed and accelerated.
This leads to the possibility of their coherent control. We then
consider this system in an optical lattice \cite{Bloch5,Warner,Kostov}, where a superfluid
phase is found to exist under general conditions.



In the case of two species condensate with a wave function
$\psi_{i}(x,t)$ for the species $i$, the coupled quasi-1D GP
equation in the presence of an external potential $V_i$, can be
written as,
\begin{widetext}
\begin{subequations}
   \begin{eqnarray}
     i\hbar \dot{\psi_{1}} &=& - \frac{\hbar^{2}}{2 m}\psi''_{1} + V_{1}(x,t)\psi_{1} + [
       g_{1}|\psi_{1}|^{2} + g_{12}|\psi_{2}|^{2} - \nu_{1}] \psi_{1}
       \label{coup.nlse1}\\
       \text{and}\;\;\;
       i\hbar \dot{\psi_{2}} &=& -
       \frac{\hbar^{2}}{2 m}\psi''_{2} + V_{2}(x,t)\psi_{2} +
       [g_{21} |\psi_{1}|^{2} + g_{2}|\psi_{2}|^{2} - \nu_{2}]
       \psi_{2}\label{coup.nlse2}.
   \end{eqnarray}
  \end{subequations}
\end{widetext}
The strength of the intra-species interactions is $g_{i}$ and
$\nu_{j}$ is the chemical potential. We assume the interspecies
interaction to be same for both the components: $g_{12}=g_{21}$;
$V_{j}$ is the trapping potential.

In the absence of any potential, the general traveling wave solutions of Eq.~(\ref{coup.nlse1}) and
(\ref{coup.nlse2}) have the following form:
\begin{subequations}
 \begin{eqnarray}
\psi_{1}(x,t) &=& \sqrt{\sigma_{01}[1 - (1-\frac{m^{2} u^{2}}{\hbar^{2}})
\sin^{2}(x - u t)]} e^{i[\chi_{1}(x,t)]},\nonumber \\ \\
\psi_{2}(x,t) &=& \sqrt{\sigma_{02}[1 -
(1-\frac{m^{2} u^{2}}{\hbar^{2}}) \cos^{2}(x - u t)]}
e^{i[\chi_{2}(x,t)]}. \nonumber \\
\end{eqnarray}
\end{subequations}
where, $\sigma_{0j}$'s are the equilibrium densities of the
atoms in the condensed phase. The phase velocity is given by,
\begin{eqnarray}\label{imag}
v_{j} = \frac{\hbar}{m} \chi'_{j} = u (1 - \frac{\sigma_{0j}}{\sigma_{j}}).
\end{eqnarray}

For these solutions to exist, it is found that the interactions need
to satisfy $g_{12}^{2} = g_{1} g_{2}$ and the background densities are
related by $g_{1} \sigma_{01} = g_{12} \sigma_{02}$. The difference
between cross phase modulation and self phase modulation leads to a
difference in chemical potentials:
\begin{eqnarray}
\nu_{1} - \nu_{2} &=& (g_{1} - g_{12})\sigma_{01}(1 +
\frac{m^{2}u^{2}}{\hbar^{2}}) \nonumber \\ &=& (g_{12} - g_{2})\sigma_{02}(1 +
\frac{m^{2}u^{2}}{\hbar^{2}}).
\end{eqnarray}
For the limiting case $u=0$, the above solutions coincide with the solutions mentioned in \cite{Warner} subjected to the zero external periodic potential.

Recently the effect of the longitudinal trap on the condensate and soliton
profile has been investigated quite intensively \cite{garcia}. In the
general scenario, the scattering length, oscillator frequencies can
be time dependent, in addition to the presence of a phenomenological
loss term \cite{atre5,xie5,alkhawaja5,utpal5}. Below we employ this
method to the sinusoidal waves in the two component scenario. As
will be seen later, this can be used for controlling the
excitations. They may be compressed or accelerated, through suitable
temporal modulations of various parameters. We consider self-similar
solutions in the oscillator trap $V_{j} = \frac{1}{2} M(t) x^{2}$,
for which the ansatz solution is of the following form,
\begin{eqnarray}
\psi_{j}(x,t) &=& \sqrt{A(t) \sigma_{j}[A(t)(x - l(t))]}
e^{i[\chi_{j}(x,t) + \phi(x,t)]}.\nonumber \\
\end{eqnarray}
Here, $\phi(x,t)$ is a density independent phase having the form
$\phi(x,t) = a(t) + b(t) x -\frac{1}{2}c(t)x^{2}$ and $l(t) =
\int_{0}^{t}{v(t') dt'}$. The sinusoidal wave, in this case, is a
propagating wave with the velocity $v(t)$ in the moving
condensate.  The consistency conditions lead to,
\begin{eqnarray}
a(t) = a_{0} - \frac{\frac{\hbar^{2}}{2 m}
-\bar{\mu}}{\hbar}\int_{0}^{t} A^{2}(t') dt'
\end{eqnarray}
where $\bar{\mu} = \mu_{j} + \lambda_{j}$. Here $\nu_{j}(t) =
\mu_{j} A^{2}(t)$ ($j = 1, 2$) and $\lambda_j$'s are constant
parameters controlling the energy of the excitations. The time
dependent wave vector $b(t) = A(t)$ and $c(t)$ can be determined
by the Ricatti type equation
\begin{eqnarray}
\hbar \frac{\partial c(t)}{\partial t} - \frac{\hbar^{2}}{m} c^{2}(t) = M(t).
\end{eqnarray}
From current conservation, amounting to solving the imaginary part of the coupled GP equations, one gets Eq.~(\ref{imag}),
with the consistency conditions:
\begin{subequations}
\begin{eqnarray}
l_{t}(t) &+& \frac{\hbar}{m}c(t) l(t) - \frac{\hbar}{m} A(t)= A(t) u \\
A(t) &=& \frac{\hbar A_{0}}{m} \exp{\int_{0}^{t}c(t') dt'}, \\
g_{j}(t) &=& \kappa_{j} A(t) \textrm{\,\,\,\, and \,\,\,\,} g_{12}(t) =
\kappa_{12} A(t).
\end{eqnarray}
\end{subequations}

\begin{figure}[h]
\includegraphics[scale=0.35]{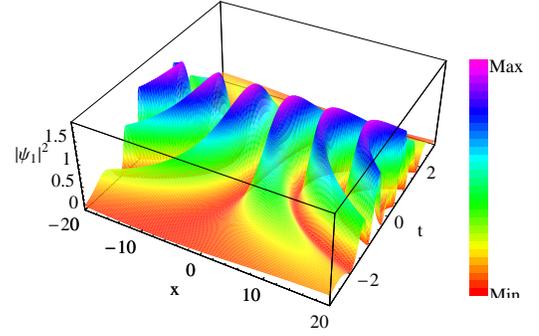}
\caption{Density distribution of the sinusoidal wave of the first
component for $^{87}Rb$ in presence of a harmonic oscillator trap
with $\kappa_{1} = 0.4, \kappa_{2} = 0.1, u = 0.8$ and $A_{0} = 0.5$.} \label{oscillator}
\end{figure}
The real part of the coupled GP equations reduces to,
\begin{widetext}
\begin{subequations}
\begin{eqnarray}
\frac{\hbar^{2}}{4 m}\sigma_{1} \sigma_{1}'' - \frac{\hbar^{2}}{8 m}
\sigma_{1}' + (\frac{1}{2}m u^{2} + \lambda_{1}) \sigma_{1}^{2} -
\kappa_{1} \sigma_{1}^{3} - \kappa_{12} \sigma_{2} \sigma_{1}^{2} -
\frac{1}{2} m u^{2} \sigma_{01}^{2} &=& 0  \\ \text{and} \;\;\;
\frac{\hbar^{2}}{4 m}\sigma_{2} \sigma_{2}'' - \frac{\hbar^{2}}{8 m}
\sigma_{2}' + (\frac{1}{2}m u^{2} + \lambda_{2}) \sigma_{2}^{2} -
\kappa_{2} \sigma_{2}^{3} - \kappa_{12} \sigma_{1} \sigma_{2}^{2} -
\frac{1}{2} m u^{2} \sigma_{02}^{2} &=& 0. \\ \nonumber
\end{eqnarray}
\end{subequations}
\end{widetext}
Consistency conditions further require $\mu=\bar{\mu} = \mu_{1} +
\lambda_{1} = \mu_{2} + \lambda_{2}$ and $\lambda_{1} - \lambda_{2} = (\kappa_{1} - \kappa_{12})\sigma_{01}(1
+ \frac{m^{2}u^{2}}{\hbar^{2}}) = (\kappa_{12} -
\kappa_{2})\sigma_{02}(1 + \frac{m^{2}u^{2}}{\hbar^{2}})$ with the
constraint $\kappa_{12}^{2} = \kappa_{1}\kappa_{2}$ and $\kappa_{1}
\sigma_{01} = \kappa_{12} \sigma_{02}$. The form of the densities have
been found to retain their earlier forms:

\begin{widetext}
\begin{subequations}
\begin{eqnarray}
\psi_{1}(x,t) &=& \sqrt{A(t) \sigma_{01}[1 - (1-\frac{m^{2}
u^{2}}{\hbar^{2}})
\sin^{2}[A(t)(x - l(t))]]} e^{i[\chi_{1}(x,t) + \phi(x,t)]} \\
\text{and} \;\;\psi_{2}(x,t) &=& \sqrt{A(t) \sigma_{02}[1 -
(1-\frac{m^{2} u^{2}}{\hbar^{2}}) \cos^{2}[A(t)(x - l(t))]]}
e^{i[\chi_{2}(x,t) + \phi(x,t)]}.
\end{eqnarray}
\end{subequations}
\end{widetext}

The non-trivial phases are now controlled by the trap:
\begin{eqnarray}
\chi_{1} = \frac{m u}{\hbar} A(t)[x - l(t)] - \tan^{-1}[\frac{m
u}{\hbar} \tan[A(t)(x - l(t))]], \nonumber \\
\end{eqnarray}
with a corresponding expression for the second component.
 The superfluid current densities in presence of the trap takes
 the form
\begin{eqnarray}
j_{1} &=& \frac{\hbar \sigma_{01}}{2 m}\big((u + A(t) - c(t)
x)(\frac{m^{2}u^{2}}{\hbar^{2}} - 1) \nonumber \\ & & \sin^{2}[A(t)(x -
l(t))]\big)
\end{eqnarray}
with a similar expression for the second component.
The flow density gets modulated by the chirped phase and as expected it depends on
the oscillator potential. Hence, by tuning the trap the current
densities can be controlled suitably.

For illustration, we first consider a trap with $M(t)= \alpha = const.$, and inter-species interactions $\kappa_{1} = 0.4$ and $\kappa_{2} = 0.1$. Mass of the $^{87}Rb$ atom is $m = 1.41 \times 10^{-25} kg.$ The equality of the SPM and XPM leads to the same background, along with the same chemical potentials for the both the components. Fig.(\ref{oscillator}) shows the traveling wave, with a time dependent velocity in the presence of the trap. In presence of oscillator, the atoms can be accelerated and suitably controlled.

It needs to be mentioned that, unlike experimentally observed localized solitons, sinusoidal solutions have infinite extent, which should be excited in a finite sized trap. In a single component BEC, periodic solutions, existing in the finite condensate, have been experimentally seen as Faraday waves \cite{engels}, which manifest when the scattering length is time dependent in a periodic manner \cite{engels}. We expect similar behavior for the sinusoidal excitations in two component Bose-Einstein condensates, since these are exact solutions.
\newline


Recently, restricted sinusoidal solutions have been found for TBEC in an optical lattice \cite{Hai5}, where the form of the optical lattice potential is taken as, $V(x) = V_{0} \cos^{2} x$, where $V_{0}$ is the amplitude of the optical lattice. The spatial co-ordinate and $V_{0}$ are scaled in the units of wavelength of incident laser light and recoil energy respetively. We find that under general conditions the following type of solutions exist:
\begin{eqnarray}
\psi_{1}(x,t) &=& \sqrt{A+B\cos^{2}(x)}
e^{i\chi_{1}(x)+i\omega_{1}t} \\ \textrm{\,\,and\,\,}
\psi_{2}(x,t) &=& \sqrt{C+D\cos^{2}(x)}
e^{i\chi_{2}(x)+i\omega_{2}t},
\end{eqnarray}
with $\omega_{j}=\frac{1}{2}+\bar{\mu_{j}}$ and $\chi_{jz}=\frac{2
c_{j}}{\rho_{j}^{2}}$ ($j=1,2)$. Here, $c_{j}'s$ are the integration
constants. Considering the scenario of independent chemical potentials
for the two species, the consistency conditions yield:
\begin{subequations}
\begin{eqnarray}
A&=&\frac{\mu_{2}g_{12}-\mu_{1}g_{2}-2V_{2}g_{12}}{g^{2}_{12}-g_{1}g_{2}},
B=\frac{V_{2}g_{12}- V_{1}g_{2}}{g^{2}_{12}-g_{1}g_{2}},\\
C&=&\frac{\mu_{1}g_{12}-\mu_{2}g_{1}+V_{2}g_{1}-V_{1}g_{12}}{g^{2}_{12}-g_{1}g_{2}},
D=\frac{V_{2}g_{1}-V_{1}g_{12}}{g^{2}_{12}-g_{1}g_{2}}, \nonumber \\
\end{eqnarray}
\end{subequations}
with $\mu_{j}=\nu_{j}+\bar{\mu}_{j}$. Dispersion only affects the
super-current through the integration constants:
$I_{1}=\frac{1}{2}AB+(\frac{1}{2}+\mu_{1})
A^{2}-g_{2}A^{3}-g_{12}(C+D)A^{2}+V_{1}A^{2}$,
where,
$I_{j}=\frac{\hbar^{2}c_{j}^{2}}{2m}$. The condensate phase for
the first component has the explicit form:
$\chi_{1}(z)=c_{1}\tan^{-1}[\frac{\sqrt{A+B}\tan(z)}{\sqrt{A}}]/\sqrt{A(A+B)}$.
Similar type of expression holds true for the second component. The difference between the solutions found here, as compared to the earlier one obtained in \cite{Hai5}, lies in the integration constants $I_{i}$. These constants acquire an additional contribution from the dispersion term in the form of $\frac{1}{2}A B$, not present in the restricted solutions found earlier. When  both the components have identical chemical potentials ($\mu_{1} = \mu_{2}$), the parameter values coincide with Ref. \cite{Hai5}.

In summary, the two component BEC is found to sustain sinusoidal excitations in a trap,
which is not possible in the single component case. It is shown that
appropriate changes in the trap and scattering length can be used to
control the BEC profile. The superfluid velocity can also be changed by
controlling the experimental parameters. We note that difference
between the ground state energy of the two components can arise because
of the XPM. The roles of both harmonic and optical
trap together is an area worthy of future investigation. It may
provide additional parameters for controlling the dynamical phase
transitions found in this system \cite{Smerzi5,manan5,altman5}. One can also study the Faraday patterns in this system with time dependent scattering length \cite{staliunas,hoefer}. The presence of the two components may affect the nature of these excitations.

\end{document}